\documentclass[12pt]{article}
\usepackage{graphicx}

\providecommand{\GeV} {\textrm{GeV}}

\providecommand{\mz} {\ensuremath{M_{\textrm{z}}}}
\providecommand{\mzsq} {\ensuremath{M^2_{\textrm{z}}}}

\providecommand{\as} {\ensuremath{\alpha_{s}}}
\providecommand{\asq} {\ensuremath{\alpha_{s}(Q^2)}}
\providecommand{\asqsq} {\ensuremath{\alpha^2_{s}(Q^2)}}
\providecommand{\asmu} {\ensuremath{\alpha_{s}(\mu^2)}}
\providecommand{\asmsq} {\ensuremath{\alpha^2_{s}(\mu^2)}}
\providecommand{\asmz} {\ensuremath{\alpha_{s}(\mzsq)}}
\providecommand{\asmzsq} {\ensuremath{\alpha^2_{s}(\mzsq)}}
\providecommand{\epem} {\ensuremath{\mathrm{e}^+\mathrm{e}^-}}

\begin{document}

\title{The Determination of the Strong Coupling Constant}

\author{G\"unther Dissertori \\
Institute for Particle Physics, ETH Zurich, Switzerland}

\maketitle
%
%%%%%%%%%%%%%%%%%%%%  ABSTRACT %%%%%%%%%%%%%%%%%%%
%
\begin{abstract}
The strong coupling constant is one of the fundamental parameters of the standard model of particle physics.
In this review I will briefly summarise the theoretical framework, within which the strong coupling constant is 
defined and how it is connected to measurable observables. 
Then I will give an historical overview of its experimental determinations and discuss the current status and world average value. 
Among the many different techniques used to determine this coupling constant in the context of quantum chromodynamics, 
I will focus in particular on a number of measurements carried out at the Large Electron Positron Collider (LEP) and the Large Hadron Collider (LHC) at CERN.Ê
\end{abstract}

\mbox{}\\[3cm]
\begin{center}
A contribution to:\\
The Standard Theory up to the Higgs discovery - 60 years of CERN \\
L. Maiani and G. Rolandi, eds.
\end{center}

\newpage
%
%%%%%%%%%%%%%%%%%%%%  SECTIONS %%%%%%%%%%%%%%%%%%%
%
\section{Introduction}
	\label{sec:intro}

The strong coupling constant, \as, is the only free parameter of the lagrangian of quantum chromodynamics (QCD),
the theory of strong interactions, if we consider the quark masses as fixed. As such, this coupling constant, or equivalently $g_s = \sqrt{4\pi\as}$, is one of the three 
fundamental coupling constants of the standard model (SM) of particle physics. It is related to the $\mathrm{SU}(3)_C$ colour part of the 
overall $\mathrm{SU}(3)_C \times \mathrm{SU}(2)_L \times \mathrm{U}(1)_Y$ gauge symmetry  of the SM.
The other two constants $g$ and $g'$ indicate the coupling strengths relevant for weak isospin and weak hypercharge,
and can be rewritten in terms of the Weinberg mixing angle $\tan\theta_\mathrm{W}=g'/g$ and the fine-structure constant 
$\alpha = e^2/(4\pi)$, where the electric charge is given by $e=g \sin\theta_\mathrm{W}$. Note that natural units are used throughout.

While typically denoted as \textit{constants}, actually all these coupling strengths vary as a function of the energy scale or momentum transfer  $Q$
of the particular process looked at, as will be discussed later. In contrast to $\alpha(Q^2)$, which increases with increasing $Q$,
the strong coupling $\asq$ decreases for increasing scale, leading to the famous property of QCD known as asymptotic freedom.
It is interesting to compare the values of these two coupling strengths at some fixed scale, such as the mass of the Z boson,
$Q\approx \mz \approx 91$ \GeV. We find that $\alpha(\mzsq) \approx 1/128$, whereas $\asmz \approx 0.12$; that is, the strong coupling
is still about 15 times larger than the fine-structure constant at energy scales much larger 
than those relevant for quark confinement into hadrons ($Q \sim 1$ \GeV).  Thus,  strong interactions are indeed strong 
compared to electroweak interactions, even at large energy scales such as those probed by CERN's past and present colliders, 
in particular the Large Electron Positron Collider (LEP) or the Large Hadron Collider (LHC).

The different energy dependence of the coupling strengths triggers the immediate question if and at which exact energy scale
these coupling constants become of equal strength, implying the onset of a possible \textit{grand unification}. Obviously, the answer to this
question also depends on the precision
at which $\alpha$ and \as\ have been determined by experiment, and it is instructive to realize
that today $\alpha(Q^2=0)\approx 1/137$ is known at an accuracy of 32 parts per billion \cite{CODATA}, whereas the relative uncertainty of
the current world average (WA) value \cite{PDG-BDS} of $\asmz = 0.1185 \pm 0.0006$  amounts to half a percent; quite an astonishing difference.

Besides the wish to improve the
accuracy of the aforementioned very high energy extrapolation, it is of general importance to know \as\ at the best possible precision, since it
enters the calculation of each and every process that involves strong interactions and thus ultimately limits the precision at which 
such processes can be predicted theoretically. As a most recent and prominent example, it is worth mentioning that the uncertainty
on \as\ gives a non-negligible contribution to the overall theoretical uncertainty on Higgs boson production at the LHC \cite{Heinemeyer:2013tqa}.
Correspondingly, this limits the studies looking for effects beyond the SM that could manifest themselves 
through deviations of the measured Higgs production cross sections from their theoretical predictions.
In the following I will indicate the experimental and theoretical difficulties that limit the precision at which we know this fundamental parameter,
but also highlight the dramatic improvements, which have been achieved during the last three decades.

%
%%%%%%%%%%%%%%%%%%%%  SECTIONS %%%%%%%%%%%%%%%%%%%
%
\section{Theoretical framework}
	\label{sec:theory}

The basic elements of QCD, including a discussion of the scale dependence of \asq\ and the 
related  structure of theoretical predictions obtained in perturbation theory, are summarized
elsewhere in this series of reviews \cite{Ellis-review}. Further extensive descriptions of the 
theoretical framework can be found in, e.g., Refs.\ \cite{PDG-BDS,Dissertori:2003pj,Ellis:1991qj}. 
Here I will only highlight  a few important aspects of perturbative QCD (pQCD), that are relevant 
for the remainder of this review.

When calculating amplitudes corresponding to Feynman graphs that involve loop diagrams, 
ultraviolet divergences are encountered. The procedure of renormalization absorbs these divergences
into a redefinition of the bare parameters and fields that appear in the lagrangian. In particular,
this leads to the renormalised or so-called \textit{running} coupling constant \asmu, a function
of the (unphysical) renormalization scale $\mu$. If $\mu$ is chosen close to the scale of the momentum
transfer $Q$ in a given process, then $\as(\mu^2 \approx Q^2)$ is indicative of the effective 
strength of the strong interaction in that process \cite{PDG-BDS}. This explains why in the literature
we often find a discussion of the running coupling constant as function of the physical scale $Q$, while
the renormalized coupling actually is a function of the unphysical scale $\mu$. This will also become 
clearer from the following discussion of the structure of perturbative predictions.

While the value of \asmu\ at a fixed scale $\mu$ can not be predicted and has to be determined from
experiment instead, its $\mu$\ dependence is given by the renormalization group equation,
\begin{equation}
	\label{eq:RGE}
	\mu^2\frac{d\as}{d\mu^2} = \beta(\as) \;\mbox{$=$}\; - b_0 \as^2 + b_1 \as^3 + \mathcal{O}(\as^4) \quad.
\end{equation}
The first two coefficients in the perturbative expansion of the so-called $\beta$-function 
of QCD are $b_0 = (33 - 2 n_f)/(12\pi)$ and $b_1 = (153 - 19 n_f)/(24\pi^2)$, where $n_f$ is
the number of ``light'' quark flavours $(m_q\ll\mu)$. Most importantly, for $n_f<17$ we have $b_0 > 0$,
which leads to a decreasing coupling strength for increasing scale (asymptotic freedom), as originally predicted by
Politzer \cite{Politzer}, Gross and Wilczek \cite{GrossWilczek}. Considering only the first term of the expansion on the right hand side of
eq.\ \ref{eq:RGE}, a solution can be written as $\asmu = \left( b_0 \ln(\mu^2/\Lambda^2) \right)^{-1}$, with $\Lambda \approx 200$ MeV
defined as the scale where \asmu\ formally diverges.
Whereas at small scales of order \GeV\ or lower
the coupling constant increases dramatically, making any perturbative approach to the solution of low-energy strong interactions
and the property of confinement meaningless, it is the property of asymptotic freedom that leads to 
an expansion parameter \as\ well below unity and thus allows perturbative methods to be applied for
the calculation of scattering processes.  

To second order, including the resummation of leading logarithms of type $\ln(\mu^2/Q^2)$,
a solution of eq.\ \ref{eq:RGE} allows to relate \as\ at one scale $\mu^2$ to that at another scale $Q^2$,
\begin{equation}
	\label{eq:RGEsol}
	\asmu = \frac{\asq}{w}\,\left(1 - \frac{b_1}{b_0} \frac{\asq}{w} \ln w  \right)\; , \quad w\,\mbox{$=$}\,1 + b_0 \asq \ln\frac{\mu^2}{Q^2}\; .	
\end{equation}
In the following we will use the resulting expansion 
\begin{equation}
	\label{eq:as-expansion}
	  \asmu\approx\asq\,\left(1 - \asq b_0 \ln\frac{\mu^2}{Q^2} + \mathcal{O}(\as^2) \right)  \quad . 
\end{equation}
This  shows that a change of scale only manifests itself as a non-leading effect in \as; in other words, a meaningful determination of the
\textit{running} coupling constant necessarily has to involve a next-to-leading order (NLO) prediction. In order to highlight this even further,
let's look at the perturbative structure, up to NLO, of some generic cross section that is proportional to \as\ at leading order (e.g., a three-jet cross section
in \epem\ annihilations at $\sqrt{s}=Q$),
\begin{equation}
	\label{eq:xsex-expansion}
	\sigma\left(\asmu,\frac{\mu^2}{Q^2}\right) = \asmu A\,+\,\asmsq \left( B + b_0 A \ln\frac{\mu^2}{Q^2} \right) \,+\, \mathcal{O}(\as^3) \quad.	
\end{equation}
The coefficients $A$ and $B$ have to be calculated for the specific process at hand. Now let us first assume that only $A$ is known
for a particular process "1", i.e., only the leading order (LO) expansion is available, $\sigma_1^{\mathrm{LO}} = \asmu A_1$. 
However, at the same LO we could equally well write this prediction as $\sigma_1^{\mathrm{LO}} = \asmu A_1 = \asq A_1$, because using
the above expansion of the coupling constant, eq.\ \ref{eq:as-expansion}, we see that the scale dependence would only appear as an NLO correction, namely
$\sigma_1^{\mathrm{LO'}} = \asq A_1 - \asqsq b_0 A_1 \ln\mu^2/Q^2$. Thus, strictly sticking to the LO expression, it is clear that an 
experimental measurement of $\sigma_1$ and its comparison to $\sigma_1^{\mathrm{LO}}$ only allows to determine some ``effective'' 
LO coupling constant $\as^{\mathrm{eff,1}}$, where it is unclear to which scale this really corresponds to. Furthermore, repeating the same procedure at LO
for some other process "2", at some different physical energy or momentum scale, would result in a measurement $\as^{\mathrm{eff,2}}$, and most likely
these two measurements of the effective coupling constant would give differing results. 

Looking again at the expression $\sigma_1^{\mathrm{LO'}} = \asq A_1 - \asqsq b_0 A_1 \ln\mu^2/Q^2$ we also see that $\sigma_1^{\mathrm{LO'}}$ strongly
depends on the unphysical scale $\mu$, since the logarithm with the explicit $\mu$ dependence already appears at NLO. Correspondingly, a determination
of $\asq$ using this prediction would result in a large uncertainty when varying the unphysical parameter $\mu$ in the fits to the measured cross section. This
procedure of $\mu$-variations, typically over a range of $0.5 < \mu/Q < 2$, has become a standard approach to estimating the possible impact
of unknown higher-order contributions. In fact, the $\mu$-dependence always appears at one order higher than the fully known perturbative expansion. 
More concretely, let's now assume that also the NLO coefficient $B$ has been calculated. Then, by plugging the expansion \ref{eq:as-expansion} into
eq.\ \ref{eq:xsex-expansion} we find 
\begin{equation}
	\label{eq:xsex-expansion2}
	\sigma\left(\asmu,\frac{\mu^2}{Q^2}\right) = \asq A\,+\,\asqsq B \,+\, \mathcal{O}(\as^3,\ln^2\frac{\mu^2}{Q^2}) \quad.	
\end{equation}
We see that there is no $\mu$-dependence up to NLO;  at this order it is thus equivalent to set $\mu=Q$ and 
to write $\sigma(\asmu,\frac{\mu^2}{Q^2}) = \sigma(\asq)$;
i.e., we can replace the dependence of the running coupling constant on the unphysical scale $\mu$ with a dependence on the physical scale $Q$ of the
process at hand. Furthermore, we see that the explicit $\mu$-dependence of the cross section prediction only appears at next-to-NLO (NNLO), 
i.e.\ suppressed by two powers of \as\ relative to the LO term. 
This leads to a  smaller uncertainty of the extracted $\asq$ value when varying $\mu$ in the fit procedure.
Finally,  the NLO expression in eq.\ \ref{eq:xsex-expansion2} leads to the first non-trivial dependence of the cross section on $\asq$ at the particular scale $Q$. 
Therefore, determinations of $\asq$ from two different processes, at possibly different values of $Q$, using the NLO predictions for the cross sections and the 
running of \as\ in order to relate the measured values to each other, should result, within uncertainties, in consistent measurements. Similarly, 
the value of \asq\ determined at NLO for some specific process can be used for predicting, at the same order, the cross section for another process at a
different physical scale. 

\begin{figure}[htbp]
   \centering
   \includegraphics[width=0.6\textwidth]{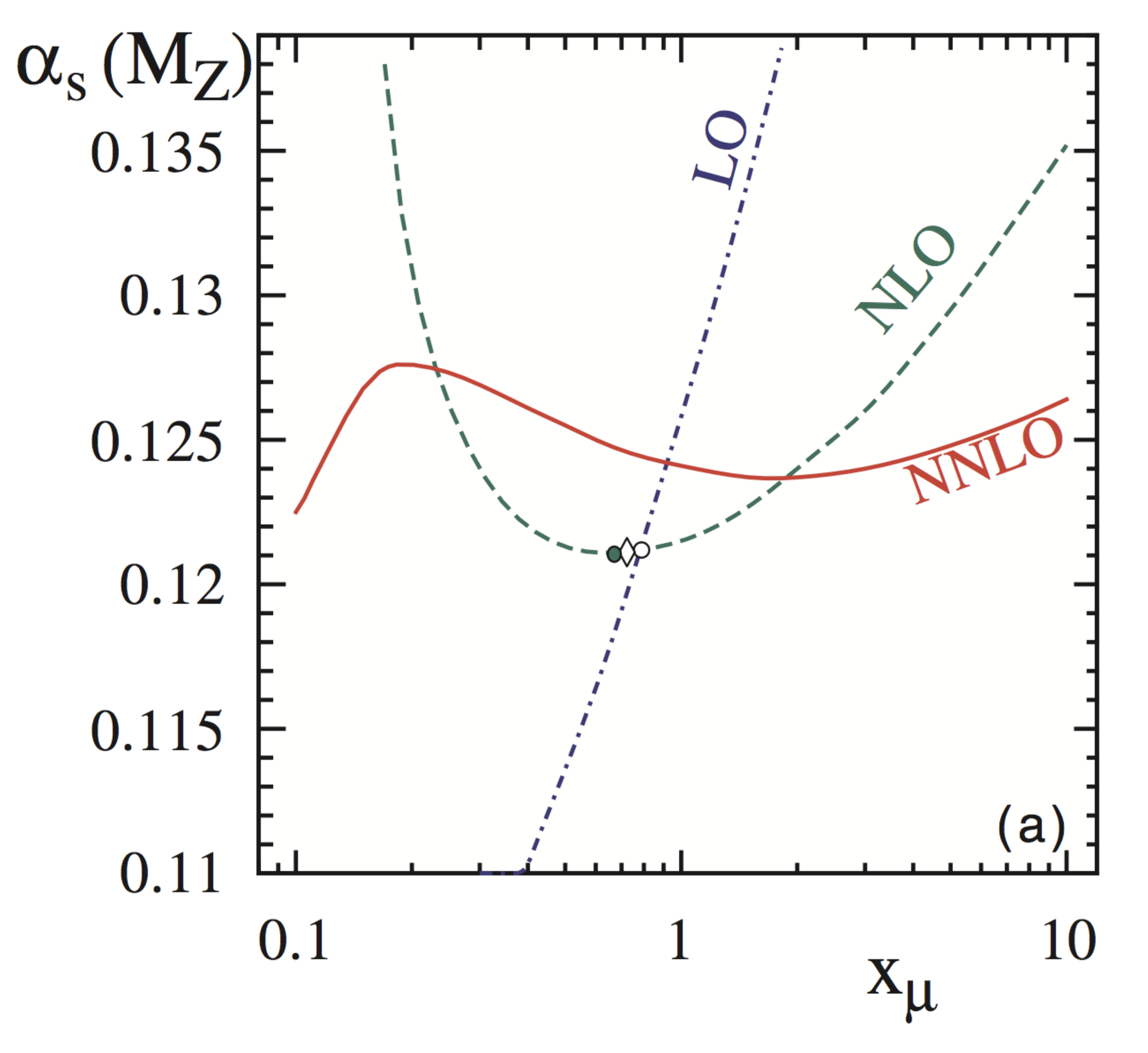} 
   \caption{\label{fig:mu-variation} \asmz\ determined from the scaled hadronic width of the Z boson, in LO, NLO and NNLO QCD, 
   	as a function of the renormalization scale $x_\mu = \mu/\mz$; taken from Ref.~\cite{Bethke:2000ai}.}
\end{figure}

The extension of this discussion to NNLO and beyond is straightforward, and easily shows that theoretical uncertainties estimated
from $\mu$-variations should decrease even further. This is nicely illustrated in Fig.\ \ref{fig:mu-variation},
where the dependence of the extracted value of \asmz\ is shown, when using the LO, NLO  and NNLO pQCD expressions for
fitting the measured hadronic width of the Z boson, normalised to its leptonic width \cite{Bethke:2000ai}. Ultimately, for an observable
known at all orders in pQCD the $\mu$-dependence would vanish completely, as it should be for a physical observable that cannot depend on 
unphysical parameters. In fact, the $\mu$-dependence of the NLO term in expression \ref{eq:xsex-expansion}  could have simply  been derived
by imposing this requirement for a physical observable and using the renormalization group equation. 

At this stage it should have become
clear, but still is worth highlighting, that the running coupling constant $\asq$ is not a physical quantity, 
but simply a parameter defined in the context of a particular theoretical framework, namely pQCD up to some order in \as. It can be 
determined experimentally in this context and used for making predictions for observables calculated within the same framework.

I would like to conclude these theoretical considerations by highlighting a further consequence of the particular scale behaviour of \as:
An uncertainty  $\delta$ on a measurement of \asq, at a scale $Q$, translates to an
  uncertainty $\delta' = (\asmzsq/\asqsq)\cdot\delta$ on \asmz; that is, $\delta' < \delta$ for $Q < \mz$.
  This enhances the impact of precise low-$Q$ measurements, such as from $\tau$ decays (c.f.\ below), in combinations performed at the \mz\ scale.

%
%%%%%%%%%%%%%%%%%%%%  SECTIONS %%%%%%%%%%%%%%%%%%%
%
\section{Observables}
	\label{sec:obs}

\begin{figure}[htbp]
   \centering
   \includegraphics[width=0.9\textwidth]{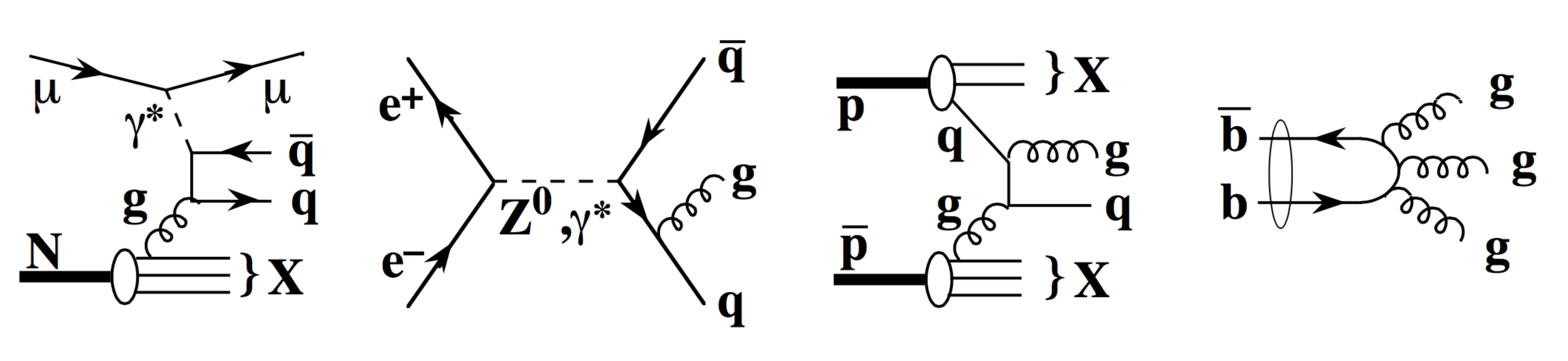} 
   \caption{\label{FeynGraphs-Bethke2004} Examples of Feynman diagrams describing hadronic final states in 
    processes which are used to measure \as; taken from Ref.~\cite{Bethke:2000ai}.}
\end{figure}

The strong coupling constant has been measured in a large variety of physics processes, using
many different observables. As depicted in Fig.\ \ref{FeynGraphs-Bethke2004}, sensitivity to the coupling
of quarks to gluons is obtained by studying, e.g., deep-inelastic lepton-nucleon scattering, 
\epem\ annihilations, hadron collisions or resonance decays. 
Since we are not able to directly measure partons (quarks or gluons), but only hadrons
 and their decay products,  a central issue is establishing a correspondence between 
observables obtained at the partonic and the hadronic level. The only theoretically sound 
correspondence is achieved by means of {\it infrared and collinear safe} (ICS) quantities (see e.g.\ Ref.\ \cite{PDG-BDS}), which
allow to obtain finite predictions at any order of perturbative QCD. ICS observables are
insensitive to the addition of a soft parton, or to the splitting of one parton into two collinear ones.
This guarantees that singularities, which appear in the infrared or collinear limits of diagrams involving real and/or virtual radiation,
cancel order by order in perturbation theory.

Generally speaking, ICS observables can be divided into different classes, depending on 
how detailed the hadronic final state is being scrutinized. As for example discussed in Ref.~\cite{PDG-BDS}, 
the simplest case of ICS observables are total cross sections. More generally, 
when measuring fully inclusive observables, the final state is not analyzed at all regarding its (topological, kinematical) structure
or its composition. Basically the relevant information consists in the rate of a process ending up
in a partonic or hadronic final state. 

In \epem\ annihilation, widely used examples
are the ratios of partial widths or branching ratios for the electroweak decay of particles into 
hadrons or leptons, such as Z or $\tau$ decays.  The strong suppression
of non-perturbative effects, ${\mathcal O}(\Lambda^4/Q^4)$, is one of the attractive features 
of such observables. However, at the same time the sensitivity to radiative QCD corrections is small, 
since here the perturbative expansion is of the type $1 + \sum_n c_n \as^n$, corresponding to, e.g., a 4\% correction,
$1+\asmz/\pi \approx 1 + 0.04$, for the scaled hadronic Z width.
In the case of $\tau$ decays not only the hadronic branching ratio is of interest, but also moments of the spectral functions of hadronic
tau decays, which sample different parts of the decay spectrum and thus provide additional information.

Other examples of fully inclusive observables, that are used for \as\ determinations, are the total top-pair production cross section in 
proton-proton collisions, the ratio of the hadronic to leptonic branching ratios of $\Upsilon$ decays, which is proportional
to $\as^3$ at LO (cf.\ the right-most diagram in Fig.\ \ref{FeynGraphs-Bethke2004}),  
or structure functions (and related sum rules) in  deep-inelastic scattering. Structure
functions are sensitive to \as\ through the Dokshitzer-Gribov-Lipatov-Altarelli-Parisi \cite{Gribov:1972ri,Lipatov:1974qm,Dokshitzer:1977sg,Altarelli:1977zs} 
evolution equations, e.g.\
$d F_2(x,Q^2)/d \ln Q^2 \propto \asq\, P_{\mathrm{qg}} \otimes g(x,Q^2)$, which depicts in a simplified manner 
the scaling violation of the $F_2$ structure function; here $x$ is the Bjorken scaling variable, $P_{\mathrm{qg}}$ 
is a so-called splitting function and $g(x,Q^2)$ is the parton distribution function (PDF) of the gluon. Such equations are used in global
PDF fits in order to relate  measurements at different $Q$ scales to each other and to fit the PDFs at a 
chosen initial scale.  We see that in such approaches the fit results for \as\ and $g(x,Q^2)$ are
strongly correlated. Similar considerations apply to the measurements of scaling violations of fragmentation functions.

\begin{figure}[htbp]
  \begin{tabular}{ll}
   \includegraphics[width=0.5\textwidth]{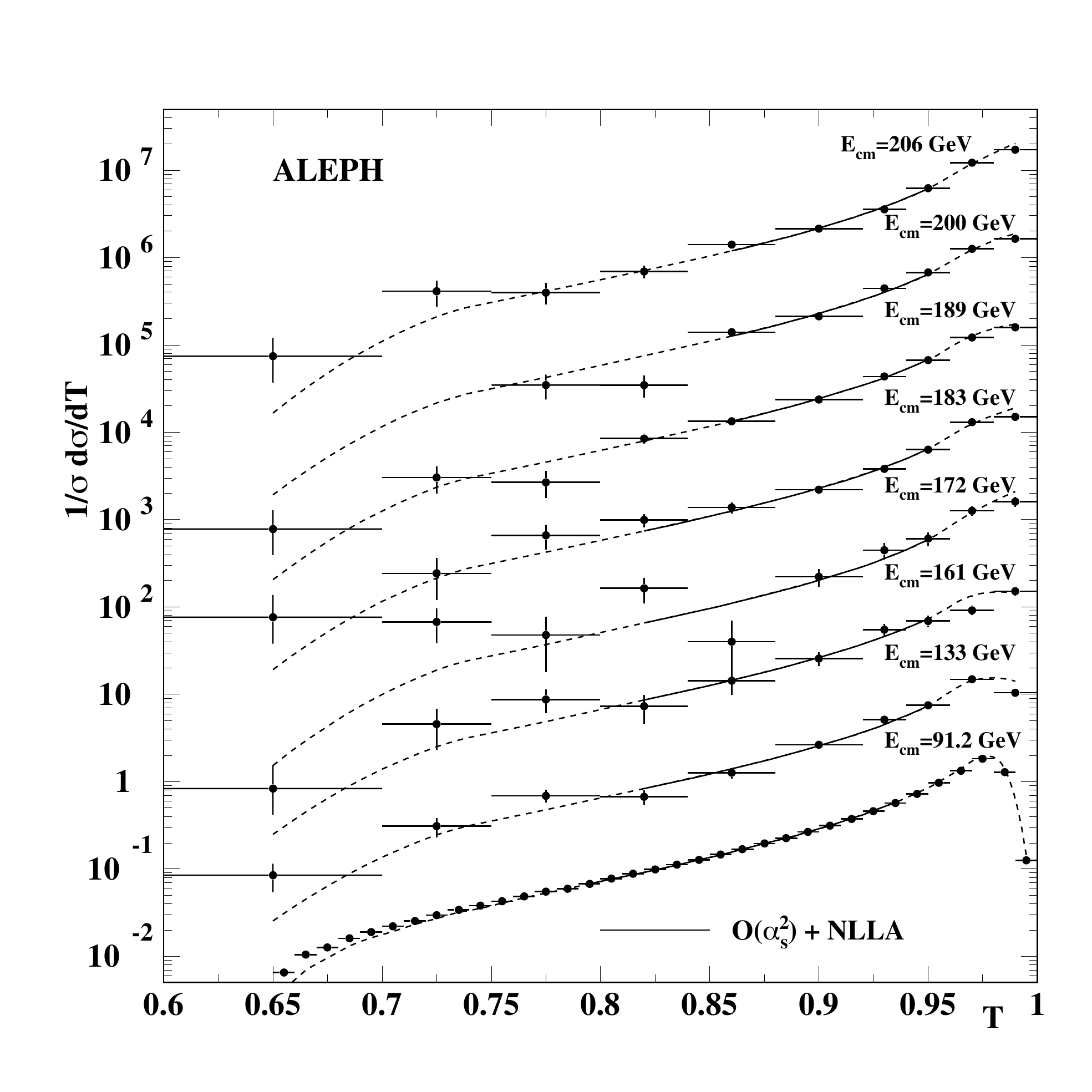} &
    \includegraphics[width=0.45\textwidth]{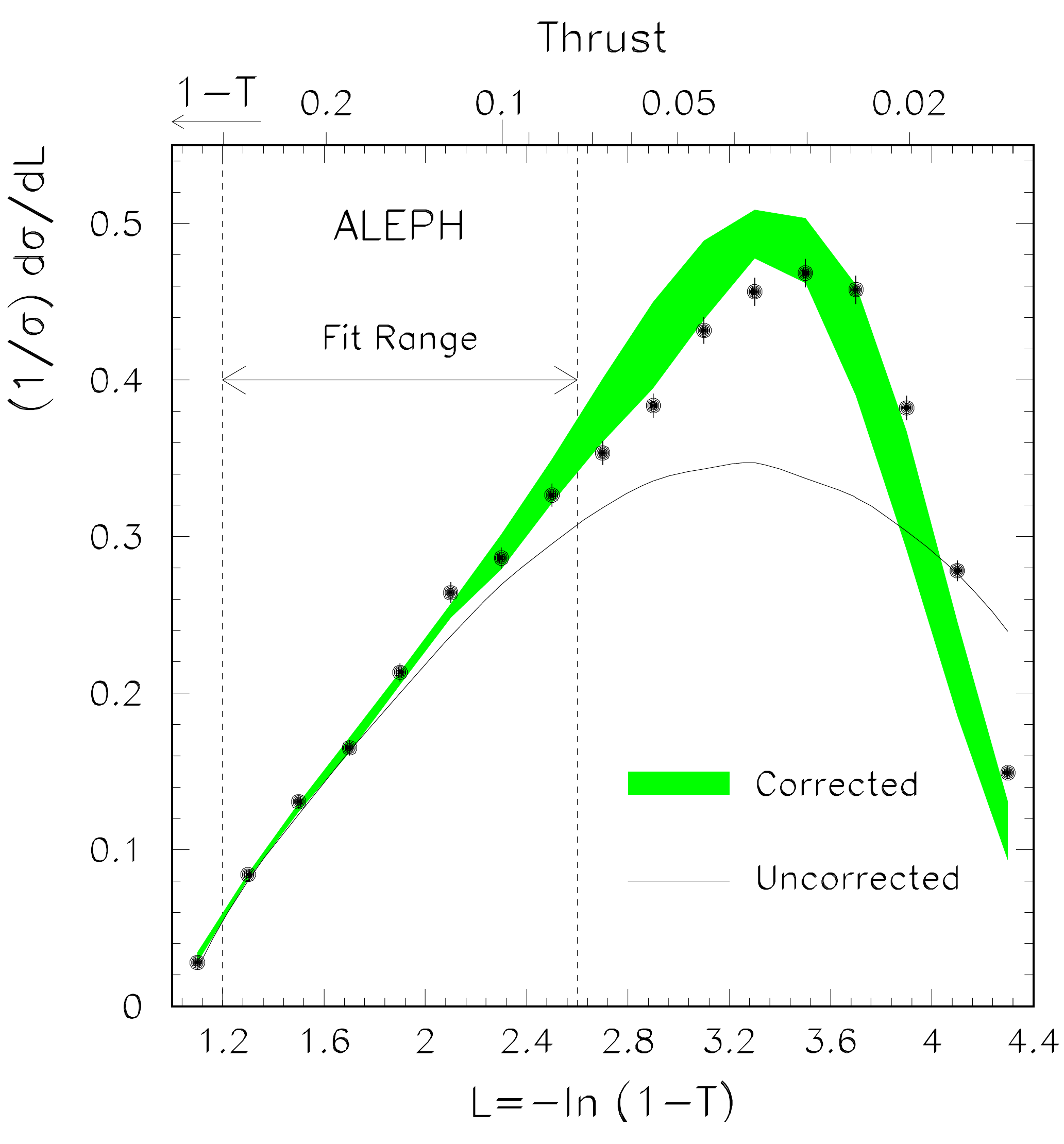} 
   \end{tabular}
   \caption{\label{fig:thrust-measurements} 
   Left: Thrust distribution measured by the ALEPH experiment at LEP for centre-of-mass energies between 91.2 and 206 GeV 
    together with QCD predictions at NLO plus next-to-leading-logarithmic approximation (NLLA). 
    The plotted distributions are scaled by arbitrary factors for presentation; taken from Ref.\ \cite{Heister:2003aj}. Right: Comparison of
     ALEPH data for the Thrust distribution to the  fitted QCD prediction (NLO+NLLA) 
     obtained at parton level (solid line) and corrected for hadronization effects (shaded band). The width of the band covers the predictions
      using different hadronization models; taken from Ref.\ \cite{Barate:1996fi}.}
\end{figure}

Compared to inclusive observables,  the particular structure, topology or composition of the hadronic final state can give
enhanced sensitivity to \as, therefore cross sections differential in one or more variables characterizing this structure are of interest.  
The obvious example is the measurement of jet cross sections and jet rates, where the probability of producing
an additional jet is directly proportional to \as\ (for a general discussion of jets and jet algorithms I refer the
reader to, e.g., Refs.~\cite{Ellis-review, PDG-BDS} and references therein). Besides jet quantities, another
class of observables, so-called \textit{event shapes}, have been
widely used, in particular for measurements in \epem\ annihilations, but also in lepton-proton and hadron collisions.
The classic example of an event shape is the 
{\it Thrust} \cite{Brandt:1964sa,Farhi:1977sg} ($T$ or $\hat\tau= 1-T$) in \epem\ annihilations, defined as
\begin{equation}
  \label{eq:thrust}
      T = \max_{\vec{n}_\tau} \frac{\sum_i |\vec{p}_i \cdot \vec{n}_\tau|}{\sum_i |\vec{p}_i|}\; ,
\end{equation}
where  $\vec{p}_i$ are the momenta of the particles or the
jets in the final-state and the maximum is obtained for the Thrust axis $\vec{n}_\tau$. In the Born limit of the 
production of a perfect  back-to-back quark-antiquark
pair the limit  $\hat\tau \rightarrow 0$ is obtained, whereas a perfectly symmetric 
many-particle configuration leads to $\hat\tau \rightarrow 1/2$. Figure \ref{fig:thrust-measurements} (left) shows an example
of measurements by the ALEPH experiment at different centre-of-mass energies. 

Besides Thrust, many other similar observables such
as C-parameter, Heavy Jet mass, Jet Broadening or the differential three-jet rate were proposed and used for \as-determinations.
They all provide a measure of the topology of an event, and typically are defined such that they take on small values
for pencil-like (back-to-back) configurations, and large values for more spherical topologies that arise from
single or multiple hard gluon radiation. This provides sensitivity to \as\ at LO in perturbation theory, with 
normalized cross sections expressed as an expansion of
the type eq.\ \ref{eq:xsex-expansion}. As discussed further below, predictions are known up to NNLO and complemented
by the all-orders resummation of logarithms of the event-shape variable (i.e., terms of the form $\as^n \ln^m\hat\tau$). 
An important aspect of event-shape variables is their enhanced sensitivity to non-perturbative effects compared
to more inclusive quantities, with power corrections of  $\sim\lambda/Q$. For \as\ determinations, 
analytical functions of these power corrections were used to complement the purely perturbative expansion, but 
the more widespread approach to correct for non-perturbative effects has been to use Monte Carlo simulations and
their hadronization models in order to calculate the event shape at the partonic and hadronic level. As can be seen
in Fig.\ \ref{fig:thrust-measurements} (right), these non-perturbative corrections can be sizeable, especially when
approaching the two-jet region of the distribution, therefore the fit range has to be chosen carefully. 
Ultimately the model dependence of such corrections leads to systematic uncertainties on the extracted \as\ values.

A completely different approach to the determination of \asmz\ consists in calculating QCD predictions on the lattice
for observables such as hadron mass splittings. From the comparison of data to the predictions, 
obtained as a function of the lattice spacing $a$ and extrapolated to $a\rightarrow 0$, first a lattice coupling is extracted
which is then converted to its perturbative counter-part \asmz. 
During the last decades there has been enormous progress in this field; indeed, the most precise 
\asmz\ determinations to date are obtained from lattice QCD, though it
is fair to say that the community still intensively discusses the validity of the very small systematic uncertainties,
claimed by some of the involved groups. A more detailed discussion of this approach can be found in the review
by Sachrajda \cite{Sachrajda-review}.

%
%%%%%%%%%%%%%%%%%%%%  SECTIONS %%%%%%%%%%%%%%%%%%%
%
\section{Brief historical overview}
	\label{sec:history}	

In the following I make an attempt to sketch some of the relevant developments that occurred during
the last few decades, without any claim of being comprehensive and of covering all types of \as\ studies 
in a balanced manner. In fact, a particular focus is put on results obtained by experiments at CERN.

\begin{figure}[htbp]
   \centering
   \includegraphics[width=0.53\textwidth,angle=90]{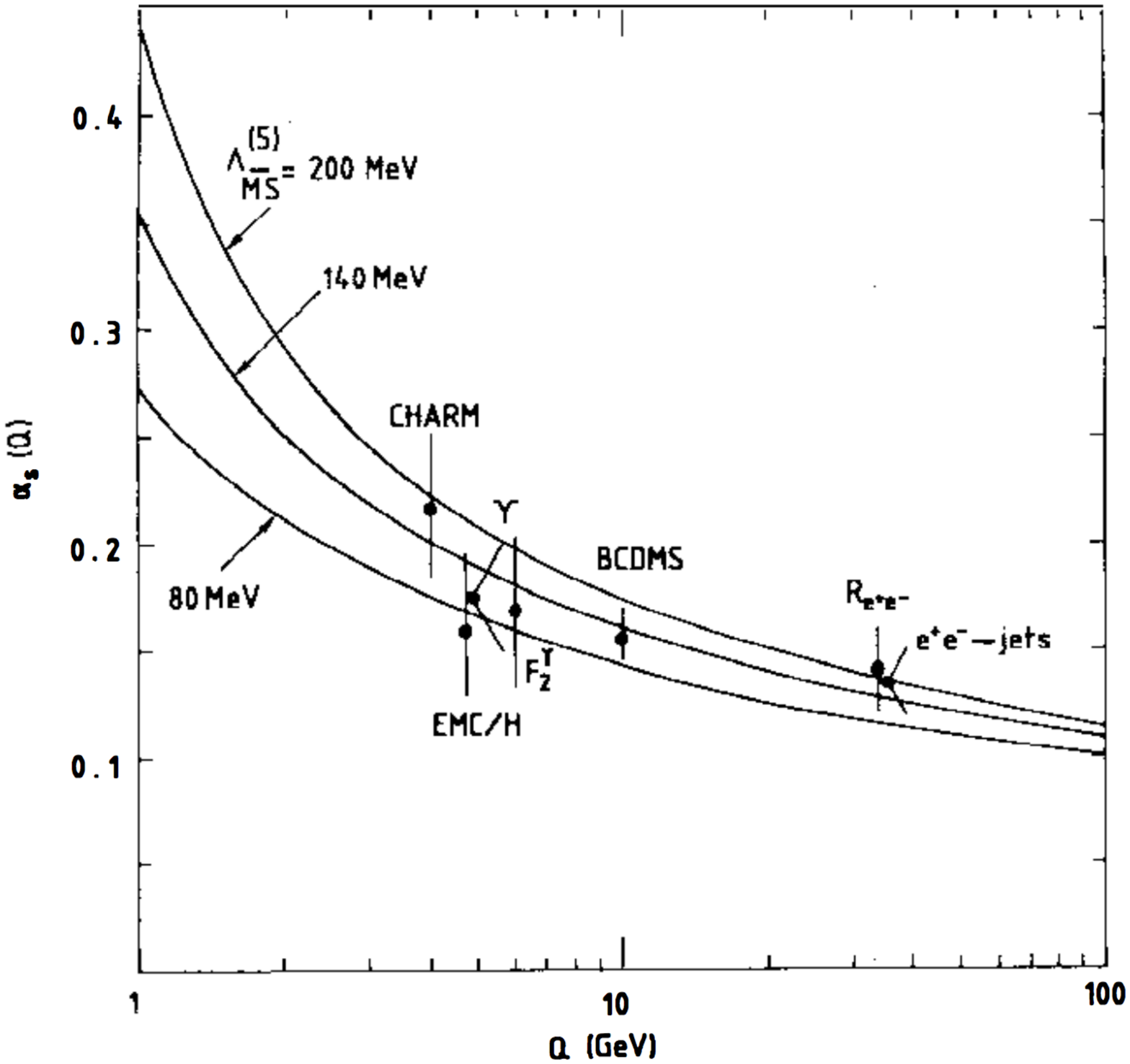} 
      \caption{\label{Altarelli-summary}Summary of the \as\ determinations by Altarelli \cite{Altarelli:1989ue} in 1989.}
\end{figure}

The first extensive overview of \as\ measurements was given by Altarelli \cite{Altarelli:1989ue} in 1989. 
In that review, he summarised measurements based on observables such as (i) the total hadronic cross section
in \epem\ annihilation (at that time known at NNLO in pQCD, i.e., up to $\as^3$) from data in the range
$7 < Q < 56$ GeV; (ii) scaling violations in deep inelastic leptoproduction with structure function data  
from BCDMS, EMC and CHARM; (iii) quarkonium decays, especially ratios of $\Upsilon$ partial widths 
($\Gamma_{\mathrm{ggg}}/\Gamma_{\mu\mu}, \Gamma_{\gamma\,\mathrm{gg}}/\Gamma_{\mathrm{ggg}}$) measured
by the CUSB, CLEO, ARGUS and Crystal Ball collaborations; and (iv) jet production, energy-energy correlations and
the photon structure function from $\gamma\gamma$ reactions, using \epem\ data collected by the 
PEP/PETRA experiments. A summary of these measurements
is shown in Fig.\ \ref{Altarelli-summary}. In an attempt to combine all
those results, obtained at $Q$ values from a few up to several tens of GeV, 
and using the QCD prediction for the running of \asq, he arrived at a prediction
for the coupling evaluated at the Z boson mass, $\as(Q\approx \mz) \approx 0.11 \pm 0.01$; 
that is, a determination of the strong coupling constant at the 10\% level. Interestingly, he concluded with the following statement:
\emph{Establishing that this prediction is experimentally true would be a very quantitative and accurate test of QCD, 
conceptually equivalent but more reasonable than trying to see the running in a given experiment}.
It is impressive to note that his prediction turned out to be nicely consistent with the current WA value \cite{PDG-BDS} 
 of $\asmz = 0.1185 \pm 0.0006$. In addition, we see that over the past 25 years the relative uncertainty has  
been reduced by a factor of 18, gauging the enormous progress made during these decades.

The year 1989 also saw the start of the LEP experiments (ALEPH, DELPHI, L3, OPAL), and the following decade was
characterised by great advances, both experimentally and theoretically, in the field of pQCD in general and \as\ measurements
in particular. Extensive overviews can be found, for instance, in Refs.\ \cite{Biebel:2001dm,Kluth:2006bw} which also discuss the application
to earlier JADE data of the developments that occurred during the LEP era. 

Event-shape observables were studied
in great detail by the LEP experiments, showing first that pQCD at NLO does not provide an adequate accuracy in order to go well below
the 10\% level in relative uncertainty on \asmz. At the same time, calculations that resum large logarithms of the event-shape variable to all orders in \as\
appeared and were used to improve the NLO prediction for a number of event-shape observables. This also triggered the 
development of a novel jet algorithm, the so-called Durham algorithm \cite{Catani:1991hj}, 
with a modified metric compared to the previously used JADE algorithm \cite{Bartel:1986ua}.
The modification of the jet metric, which defines the distance in phase-space between two particles that might or might not be combined
into a new pseudo-particle, was motivated by the fact that the pQCD predictions for jet rates and the differential 3-jet rate, based on
the JADE metric, did not show the exponentiation behaviour of the large logarithms \cite{Brown:1990nm}, whereas using the Durham metric led to
exponentiation and ultimately to improved resummation predictions. Note that this Durham algorithm became the standard algorithm for
jet finding at LEP, and was at the basis for later developments of iterative recombination algorithms, nowadays widely used 
at the LHC (cf.\ Ref.\ \cite{PDG-BDS} and references therein). As a consequence of the combined NLO+resummed predictions,
corrected for non-perturbative hadronization effects using phenomenological Monte Carlo (MC) models, the relative uncertainty of \as\ 
was reduced to the $\sim5\%$ level, still dominated by theoretical uncertainties due to missing higher orders and estimated
from variations of the renormalisation scale. Attempts to replace the MC models by analytical power corrections \cite{Dokshitzer:1995zt} of order $\lambda/Q$ did not 
lead to substantially different results. It became clear that only a complete NNLO calculation for jet rates and event shapes might
lead to a significant reduction of uncertainties. Indeed, such a calculation \cite{GehrmannDeRidder:2007bj,GehrmannDeRidder:2007hr} 
 appeared  after the end of LEP, and its first applications \cite{Dissertori:2007xa,Dissertori:2009ik} to 
the 3-jet rate and to event shapes, including next-to-leading log resummations, resulted in more precise \as\ measurements at the 2-3\% precision level.

Most of the aforementioned determinations gave \asmz\ values in a range of, very roughly speaking, $0.117$ - $0.125$. However,
as summarized in Ref.\ \cite{PDG-BDS}, more recent re-analyses of the Thrust distribution, based on novel developments in soft-collinear effective field theory, resummation
at next-to-next-to-next-to-leading logarithmic accuracy, and analytic calculations of non-perturbative effects, resulted in values as low
as $0.1131$, while at the same time claiming very small uncertainties at the 2\% level. Thus, further work will be required to understand
this spread of results from jet and event-shape observables, which covers a larger range than most of the individually quoted uncertainties.

In terms of inclusive observables, the LEP experiments quoted precise \as\ measurements by using the total hadronic cross section
(or equivalently, the hadronic width of the Z boson), as well as by analyzing hadronic $\tau$ decays. Contrary to event shapes, NNLO predictions
for these observables were already available in the nineties, leading to rather small renormalisation scale uncertainties. By now, they are
even known to N$^3$LO accuracy.  This implies an almost negligible theoretical uncertainty in the case of the hadronic Z decay width; for instance,
when included in a global fit \cite{Flacher:2008zq} to electroweak precision data a value of  $\asmz=0.1193 \pm 0.0028$ is found, where the dominant part of the uncertainty
is of statistical nature. 

Naively speaking, similarly precise results might not have been expected from the analyses of hadronic $\tau$ decays, since here
the relevant scale is the $\tau$ mass, close to the scale where pQCD is supposed to break down. Thus, non-perturbative
effects and missing higher order contributions should significantly impact the attainable precision. However, it was realised that measuring
different moments of the $\tau$ spectral function allows to determine $\as(M^2_\tau)$ and to constrain non-perturbative power-suppressed contributions 
at the same time. Several methods were developed to estimate missing higher-order terms, beyond NNLO and N$^3$LO, such as so-called
contour-improved perturbative expansions, claiming very small scale uncertainties. It is worth noting \cite{PDG-BDS} that these methods
are still matter of intense discussions, in particular since some of the results obtained by different groups differ by several standard deviations in terms
of the quoted uncertainties. In an attempt to combine all these results and to take into account the observed spread, Ref.\ \cite{PDG-BDS} 
quotes $\asmz=0.1197 \pm 0.0016$. Thus, somewhat surprisingly, the study of $\tau$ decays results in one of the most precise \as\ determinations,
basically at the level of 1\% relative uncertainty. It should be emphasized that the already precise measurements, obtained at the scale of the $\tau$ mass,
turn into this even more precise result at the Z mass, because of the running of \as\ as discussed at the end of Section \ref{sec:theory}.

Many of the developments of the LEP area, in the field of event shapes and jet observables, were also applied to HERA data of deep inelastic 
electron--proton scattering (DIS). Although here the pQCD predictions are only known up to NLO approximation so far, and the \as\ extraction from
jet final states is somewhat more complicated because of the additional implicit \as\ dependence of the PDFs, it is impressive to see that a combination \cite{Glasman:2007sm}
based on precise HERA data of inclusive jet cross sections in neutral current DIS at high $Q^2$ results in $\asmz=0.1198\pm0.0032$, which
includes a theoretical uncertainty of $\pm0.0026$. These HERA measurements also allow covering
a large range of $Q^2$ values and thus probing directly the running of \as. 

More inclusive DIS observables, in particular structure functions and
their scaling violations as discussed in Section \ref{sec:obs}, have been used in global PDF fits based on NNLO pQCD and resulted in
smaller relative uncertainties, even at the 1\% level as quoted by some groups. However, quite similarly to the case of event shapes and $\tau$ observables,
also here a spread of \asmz\ values 
(roughly covering a range of $0.113$ \cite{Alekhin:2012ig} to $0.117$ \cite{Martin:2009bu,Ball:2011us}) 
is observed \cite{PDG-BDS} that is larger than some of the 
individually quoted uncertainties. Two remarks are in place here: (i) these differences are still matter of intense discussions among the
various groups performing global PDF fits,  and (ii) it is kind of a tradition that \as\ determinations from DIS and global PDF fits result
in smaller values than those obtained from \epem\ annihilations, without understanding  the origin of this apparent bias.

Jet observables at hadron colliders, such as the inclusive jet cross section as a function of jet transverse momentum or
invariant multi-jet masses, jet angular correlations or jet rates, are only known to NLO approximation so far. Furthermore, 
important systematic uncertainties due to the jet energy scale and choice of PDF set are expected to limit the attainable precision,
and similarly to the case of DIS, the intrinsic \as\ dependence of the PDFs has to be carefully taken into account in any
\as\ fit. As discussed in Ref.\ \cite{PDG-BDS}, first measurements at the Tevatron and at the LHC gave results consistent with the
WA value and with relative uncertainties in the range of 4 to 8\%. However, very important developments have taken place
at the LHC recently, as e.g.\ summarised in Ref.\ \cite{Rojo:2014kta}. First, in both the ATLAS and CMS experiments 
the jet energy scale uncertainty is now known at an impressive level of 
about 1-2\% for jets in the $\sim100$ GeV range \cite{Aad:2014bia,Chatrchyan:2011ds}. Since jet cross sections
are steeply falling functions of jet momentum this has an enormous impact on the finally attainable precision. Furthermore, ratios of
observables, such as the ratio of the 3-jet over the 2-jet rate, allow to eliminate this systematic uncertainty to a large extend, as shown
in Refs.\ \cite{Chatrchyan:2013txa,ATLAS3jet}. Finally, NNLO calculations for the inclusive jet cross section appear to be around the corner \cite{Pires:2014rxa}, 
which will further boost the importance of such measurements. 

In fact, recently the first \as\ determination \cite{Chatrchyan:2013haa}
at a hadron collider, using pQCD at NNLO, 
has been published. However, here an inclusive quantity, namely the top-pair production cross section, has been successfully exploited thanks to
its strong sensitivity to both \as\ and the top quark mass. Fixing the latter to its WA value allowed the CMS collaboration to extract \asmz\ at an
impressive relative precision of $\sim3$\%, also thanks to the remarkable experimental precision (4\%) 
of the top cross section measurement \cite{Chatrchyan:2012bra} that served as input. Because of this recent progress, and because of the large $Q^2$ 
range covered by the measurements at the LHC, the running of the strong coupling constant is now being precisely studied over an unprecedented energy range.

As mentioned at the end of Section \ref{sec:obs}, a discussion of \as\ determinations using lattice QCD can be found in a separate review \cite{Sachrajda-review} in
this volume. For completeness it should be stated here that this very complementary approach, compared to the measurements described above,
results in the world's most precise \asmz\ determinations to date, with some of the analyses quoting relative uncertainties at the 0.5\% level (cf.\ Ref.\ \cite{PDG-BDS}).
However, the community is having  intense discussions about the validity of these
apparently rather optimistic estimates of systematic uncertainties. In any case, the lattice results dominate the current WA value: 
not including them in the averaging procedure  results in $\asmz = 0.1183 \pm 0.0012$ \cite{PDG-BDS}, i.e., the uncertainty doubles.

This historical overview can not be concluded without a brief discussion of the general issue of combining \as\ determinations. As 
discussed in Ref.\ \cite{PDG-BDS}, this is a non-trivial exercise. Since most of the individual measurements are dominated by systematic
uncertainties, which cannot be expected to follow a normal distribution, and since very often the correlations among these uncertainties
are not very well known, simple averaging methods as applicable to measurements with statistical errors only might not be appropriate.
In 1995 Schmelling \cite{Schmelling:1994pz} proposed a method for estimating such unknown correlations, which rescales individual
uncertainties according to the assumption that the normalized $\chi^2$ of the combination should be 1. This method is also used for the current
WA determination \cite{PDG-BDS}.
Furthermore, there is a certain arbitrariness in the choice of results included in the average. Finally, as mentioned earlier, often
\as\ determinations based on the same observable but using different methods give results that differ by a larger amount than would be
expected from the individually quoted uncertainties, rendering the estimate of the combined uncertainty a difficult exercise. 

\begin{figure}[thb]
\centering\small
 \begin{tabular}{|l|c|c|}
 \hline
  & & \\[-0.3cm]
  & year & World average \asmz\ \\ 
 \hline
   & & \\[-0.3cm]
 Altarelli \cite{Altarelli:1989ue}   & 1989 &  $ 0.11 \pm 0.01$ \\
 Hinchcliffe \cite{Hikasa:1992je} (PDG) & 1992  &  $0.1134 \pm 0.0035$ \\
 Hinchcliffe \cite{Montanet:1994xu} (PDG) & 1995 & $0.118 \pm 0.003$ \\
 Schmelling \cite{Schmelling:1996wm}           & 1997  & $0.118 \pm 0.003$ \\
 Bethke \cite{Bethke:2000ai}		& 2000 &     $0.1184 \pm 0.0031$ \\
 Bethke \cite{Bethke:2006ac}	         & 2006 &    $0.1189 \pm 0.0010$ \\
 Bethke, Dissertori, Salam (PDG) \cite{PDG-BDS} & 2013 & $0.1185 \pm 0.0006$ \\
 \hline
 \end{tabular}
 \includegraphics[width=0.67\textwidth]{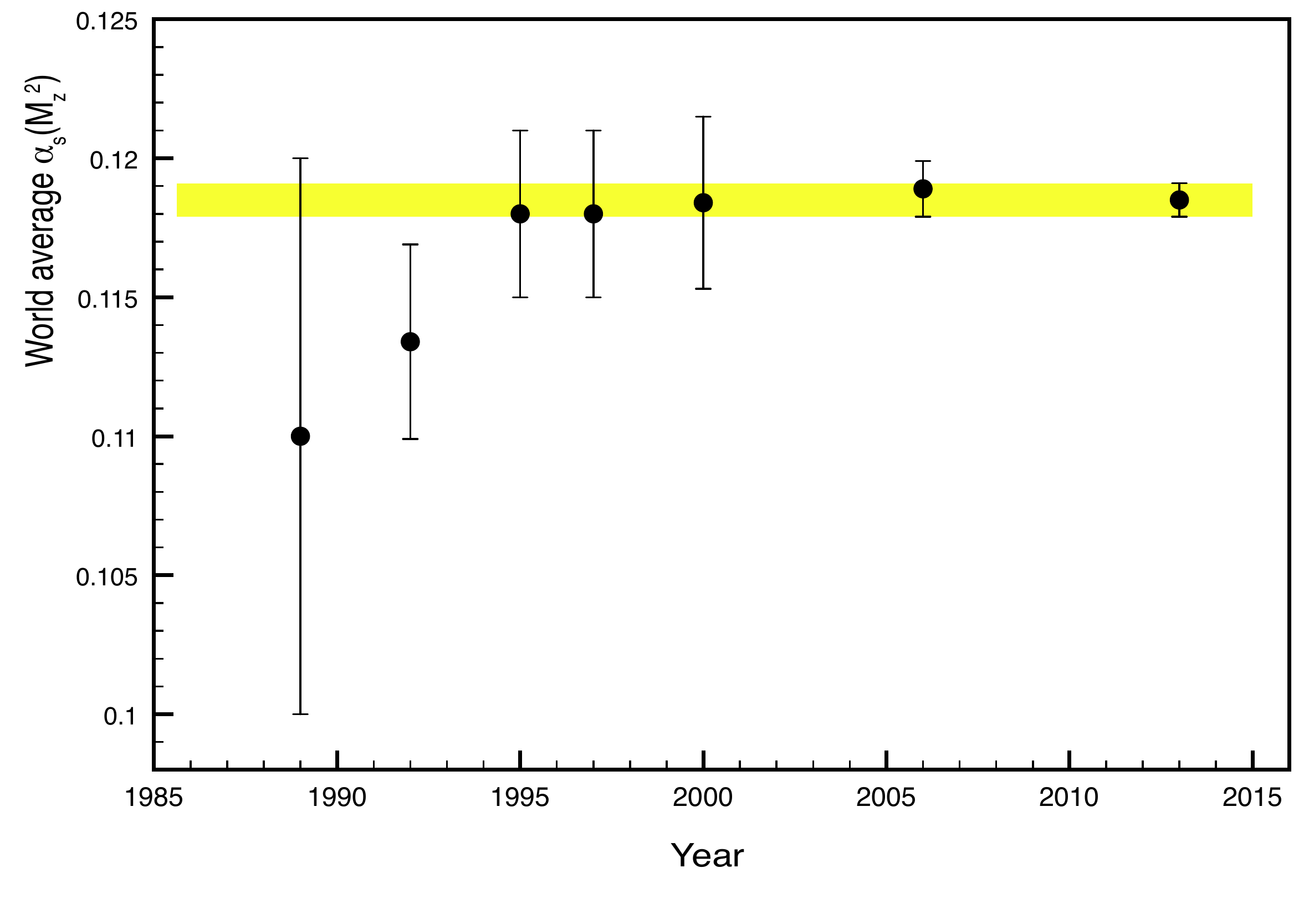}
   \caption{A selection of world average values for \asmz\ as a function of time; the yellow band indicates the current world average value \cite{PDG-BDS} of
   $\asmz = 0.1185\pm0.0006$.}
   \label{fig:asWA}
\end{figure}

\begin{figure}[htbp]
   \begin{tabular}{ll}
   \includegraphics[width=0.28\textwidth]{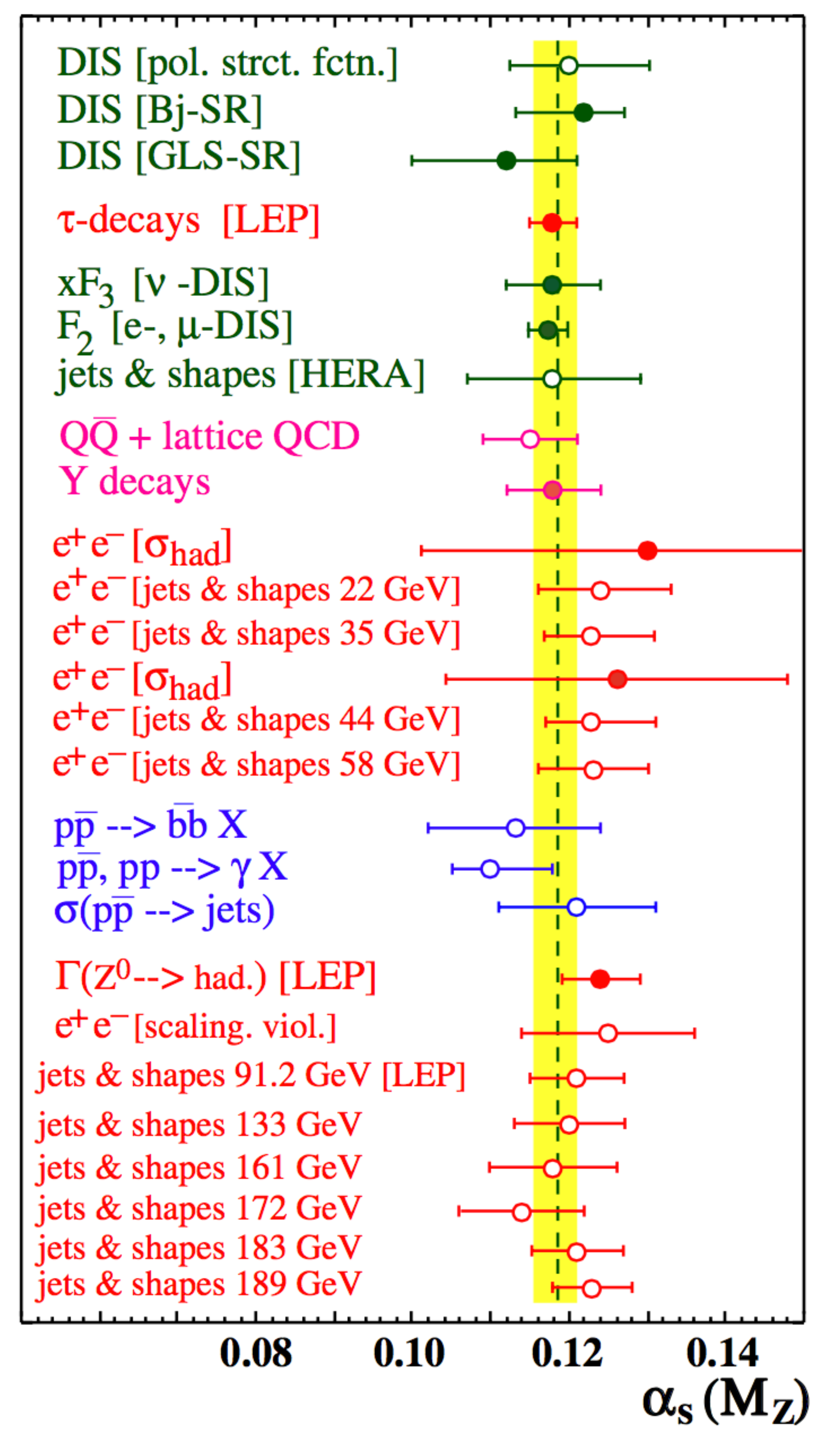} &
   \includegraphics[width=0.7\textwidth]{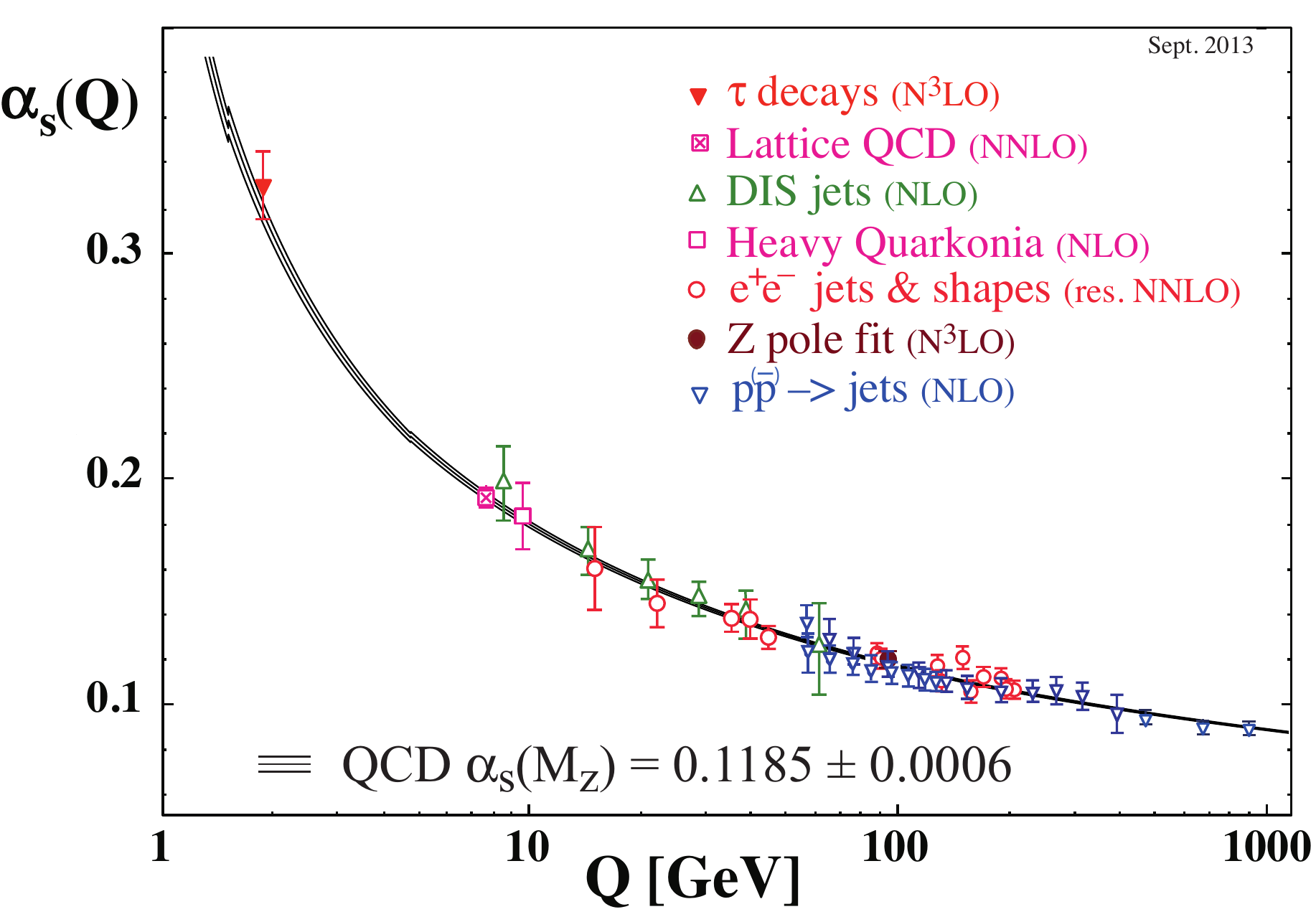} 
      \end{tabular}
   \caption{Left: List of individual \asmz\ measurements and their comparison to the world average from Ref.\ \cite{Bethke:2000ai} in 2000; Right:
   current status of the running of \as, as summarised in Ref.\ \cite{PDG-BDS}.}
   \label{fig:asq-2013}
\end{figure}

Throughout these years, several individuals and/or groups have compiled the available \as\ measurements and combined them into
a single value. The earliest attempt by Altarelli has already been discussed above. During the nineties, the reference in terms of \asmz\ 
was established by the PDG, in particular thanks to the PDG review on QCD by Hinchcliffe (see, e.g., Refs.\ \cite{Hikasa:1992je,Montanet:1994xu}).
An independent estimate of the WA value was published by Schmelling \cite{Schmelling:1996wm} in 1997, based on his proposal for handling
unknown correlations. Then, during the first decade of this century, Bethke \cite{Bethke:2000ai,Bethke:2004uy,Bethke:2006ac} provided a number of comprehensive studies,
that established the de-facto WA value, despite the PDG still publishing an independent combination. Since a few years this
situation has been resolved, with Bethke now being co-author (together with Dissertori and Salam) of the PDG review on QCD that also contains
the WA determination of \as. Figure \ref{fig:asWA} displays this, most likely incomplete, collection of WA results as a function
of time, nicely showing the impressive progress made throughout the last decades.
Finally, Fig.\ \ref{fig:asq-2013} presents an example \cite{Bethke:2000ai} of inputs to the averaging procedure  and the current experimental status of the 
running of \as, showing excellent agreement with the theoretical expectation.

%
%%%%%%%%%%%%%%%%%%%%  SECTIONS %%%%%%%%%%%%%%%%%%%
%
\section{Conclusions}
	\label{sec:conclusions}

The strong coupling constant is one of the fundamental parameters of the standard model of particle physics.
In this article I have reviewed the theoretical and  experimental developments that have led
to a precise knowledge of this important parameter, representing a cornerstone  
in our understanding of the strong interactions sector of the standard model.

%
%%%%%%%%%%%%%%%%%%%%  Thanks %%%%%%%%%%%%%%%%%%%
%
\section{Acknowledgements}
	\label{sec:thanks}

I would like to thank G.\ Rolandi and L.\ Maiani for inviting me to contribute to this collection of essays
on the Standard Theory. I would also like to thank  S.\ Bethke and G.\ Salam for many interesting discussions on the topic of \as\ and for their comments on
the manuscript. Finally, my thanks go to R. Miquel for providing me the coloured version of Fig.\ \ref{fig:thrust-measurements} (right).

%
%%%%%%%%%%%%%%%%%%%%  BIB %%%%%%%%%%%%%%%%%%%
%


\begin{thebibliography}{9}        % for non BIBTeX users

\bibitem{CODATA} $\tt{http://physics.nist.gov/cuu/Constants/index.html}$

\bibitem{PDG-BDS} S.~Bethke, G.~Dissertori and G.~Salam, \textit{Quantum Chromodynamics}, in: 
	      K.A.\ Olive \textit{et al.} (Particle Data Group), Chin.\ Phys.\ \textbf{C38} (2014) 090001.

%\cite{Heinemeyer:2013tqa}
\bibitem{Heinemeyer:2013tqa}
  S.~Heinemeyer {\it et al.}  [LHC Higgs Cross Section Working Group Collaboration],
  %``Handbook of LHC Higgs Cross Sections: 3. Higgs Properties,''
  arXiv:1307.1347 [hep-ph].

\bibitem{Ellis-review} 
R.~K.~Ellis, \textit{Quantum Chromodynamics and Deep Inelastic Scattering}, contribution to 
``Standard Theory: Essays in the 60th Anniversary of CERN'', World Scientific Publishing, eds. L.~Maiani and G.~Rolandi, 2015.
  
%\cite{Dissertori:2003pj}
\bibitem{Dissertori:2003pj}
  G.~Dissertori, I.~G.~Knowles and M.~Schmelling,
  \textit{Quantum Chromodynamics: High energy experiments and theory}, International series of monographs on physics, vol.\ 115, Oxford Univ. Press.

%\cite{Ellis:1991qj}
\bibitem{Ellis:1991qj}
  R.~K.~Ellis, W.~J.~Stirling and B.~R.~Webber, \textit{QCD and collider physics},
  Camb.\ Monogr.\ Part.\ Phys.\ Nucl.\ Phys.\ Cosmol.\  {\bf 8} (1996) 1.

\bibitem{Politzer} 
H.~D.~Politzer, \textit{Reliable Perturbative Results for Strong Interactions?}, Phys.\ Rev.\ Lett.\
{\bf 30} (1973) 1346; doi:10.1103/PhysRevLett.30.1346.    

\bibitem{GrossWilczek} 
D.~Gross and F.~Wilczek, \textit{Asymptotically Free Gauge Theories. 1}, Phys.\ Rev.\ {\bf D8} (1973) 3633; doi:10.1103/PhysRevD.8.3633;\\
D.~Gross and F.~Wilczek,  \textit{Asymptotically Free Gauge Theories. 2}, Phys.\ Rev.\ {\bf D9} (1980) 980; doi: 10.1103/PhysRevD.9.980.

%\cite{Bethke:2000ai}
\bibitem{Bethke:2000ai}
  S.~Bethke,
  %``Determination of the QCD coupling $\alpha_s$,''
  J.\ Phys.\ G {\bf 26} (2000) R27
  [hep-ex/0004021].
  %%CITATION = HEP-EX/0004021;%%


%\cite{Gribov:1972ri}
\bibitem{Gribov:1972ri}
  V.~N.~Gribov and L.~N.~Lipatov,
  %``Deep inelastic e p scattering in perturbation theory,''
  Sov.\ J.\ Nucl.\ Phys.\  {\bf 15} (1972) 438
   [Yad.\ Fiz.\  {\bf 15} (1972) 781].
  %%CITATION = SJNCA,15,438;%%
  %3039 citations counted in INSPIRE as of 13 Jun 2015
  
  
  %\cite{Lipatov:1974qm}
\bibitem{Lipatov:1974qm}
  L.~N.~Lipatov,
  %``The parton model and perturbation theory,''
  Sov.\ J.\ Nucl.\ Phys.\  {\bf 20} (1975) 94
   [Yad.\ Fiz.\  {\bf 20} (1974) 181].
  %%CITATION = SJNCA,20,94;%%
  %1138 citations counted in INSPIRE as of 13 Jun 2015
  
  %\cite{Dokshitzer:1977sg}
\bibitem{Dokshitzer:1977sg}
  Y.~L.~Dokshitzer,
  %``Calculation of the Structure Functions for Deep Inelastic Scattering and e+ e- Annihilation by Perturbation Theory in Quantum Chromodynamics.,''
  Sov.\ Phys.\ JETP {\bf 46} (1977) 641
   [Zh.\ Eksp.\ Teor.\ Fiz.\  {\bf 73} (1977) 1216].
  %%CITATION = SPHJA,46,641;%%
  %2682 citations counted in INSPIRE as of 13 juin 2015
  
  %\cite{Altarelli:1977zs}
\bibitem{Altarelli:1977zs}
  G.~Altarelli and G.~Parisi,
  %``Asymptotic Freedom in Parton Language,''
  Nucl.\ Phys.\ B {\bf 126} (1977) 298.
  %%CITATION = NUPHA,B126,298;%%
  %5300 citations counted in INSPIRE as of 13 Jun 2015
  
  

%\cite{Brandt:1964sa}
\bibitem{Brandt:1964sa}
  S.~Brandt, C.~Peyrou, R.~Sosnowski and A.~Wroblewski,
  %``The Principal axis of jets. An Attempt to analyze high-energy collisions as two-body processes,''
  Phys.\ Lett.\  {\bf 12} (1964) 57.
  %%CITATION = PHLTA,12,57;%%
  
 %\cite{Farhi:1977sg}
\bibitem{Farhi:1977sg}
  E.~Farhi,
  %``A QCD Test for Jets,''
  Phys.\ Rev.\ Lett.\  {\bf 39} (1977) 1587.
  %%CITATION = PRLTA,39,1587;%%
  
  
  %\cite{Heister:2003aj}
\bibitem{Heister:2003aj}
  A.~Heister {\it et al.}  [ALEPH Collaboration],
  %``Studies of QCD at e+ e- centre-of-mass energies between 91-GeV and 209-GeV,''
  Eur.\ Phys.\ J.\ C {\bf 35} (2004) 457.
  %%CITATION = EPHJA,C35,457;%%
  
  %\cite{Barate:1996fi}
\bibitem{Barate:1996fi}
  R.~Barate {\it et al.}  [ALEPH Collaboration],
  %``Studies of quantum chromodynamics with the ALEPH detector,''
  Phys.\ Rept.\  {\bf 294} (1998) 1.
  %%CITATION = PRPLC,294,1;%%
  
\bibitem{Sachrajda-review} 
C.~Sachrajda, \textit{Results from Lattice QCD}, contribution to 
``Standard Theory: Essays in the 60th Anniversary of CERN'', World Scientific Publishing, eds. L.~Maiani and G.~Rolandi, 2015.

%\cite{Altarelli:1989ue}
\bibitem{Altarelli:1989ue}
  G.~Altarelli,
  %``Experimental Tests of Perturbative {QCD},''
  Ann.\ Rev.\ Nucl.\ Part.\ Sci.\  {\bf 39} (1989) 357.
  %%CITATION = ARNUA,39,357;%%
  %103 citations counted in INSPIRE as of 04 Jun 2015


%\cite{Kluth:2006bw}
\bibitem{Kluth:2006bw}
  S.~Kluth,
  %``Tests of Quantum Chromo Dynamics at e+ e- Colliders,''
  Rept.\ Prog.\ Phys.\  {\bf 69} (2006) 1771
  [hep-ex/0603011].
  %%CITATION = HEP-EX/0603011;%%
  %59 citations counted in INSPIRE as of 04 juin 2015
  
 %\cite{Biebel:2001dm}
\bibitem{Biebel:2001dm}
  O.~Biebel,
  %``Experimental tests of the strong interaction and its energy dependence in electron positron annihilation,''
  Phys.\ Rept.\  {\bf 340} (2001) 165.
  %%CITATION = PRPLC,340,165;%%
  %43 citations counted in INSPIRE as of 04 Jun 2015

%\cite{Catani:1991hj}
\bibitem{Catani:1991hj}
  S.~Catani, Y.~L.~Dokshitzer, M.~Olsson, G.~Turnock and B.~R.~Webber,
  %``New clustering algorithm for multi - jet cross-sections in e+ e- annihilation,''
  Phys.\ Lett.\ B {\bf 269} (1991) 432.
  %%CITATION = PHLTA,B269,432;%%
  %881 citations counted in INSPIRE as of 04 juin 2015

%\cite{Bartel:1986ua}
\bibitem{Bartel:1986ua}
  W.~Bartel {\it et al.}  [JADE Collaboration],
  %``Experimental Studies on Multi-Jet Production in e+ e- Annihilation at PETRA Energies,''
  Z.\ Phys.\ C {\bf 33} (1986) 23.
  %%CITATION = ZEPYA,C33,23;%%
  %830 citations counted in INSPIRE as of 04 Jun 2015
  
 %\cite{Brown:1990nm}
\bibitem{Brown:1990nm}
  N.~Brown and W.~J.~Stirling,
  %``Jet cross-sections at leading double logarithm in e+ e- annihilation,''
  Phys.\ Lett.\ B {\bf 252} (1990) 657.
  %%CITATION = PHLTA,B252,657;%%
  %189 citations counted in INSPIRE as of 04 juin 2015

%\cite{Dokshitzer:1995zt}
\bibitem{Dokshitzer:1995zt}
  Y.~L.~Dokshitzer and B.~R.~Webber,
  %``Calculation of power corrections to hadronic event shapes,''
  Phys.\ Lett.\ B {\bf 352} (1995) 451
  [hep-ph/9504219].
  %%CITATION = HEP-PH/9504219;%%
  %339 citations counted in INSPIRE as of 04 juin 2015


%\cite{GehrmannDe Ridder:2007bj}
\bibitem{GehrmannDeRidder:2007bj}
  A.~Gehrmann-De Ridder, T.~Gehrmann, E.~W.~N.~Glover and G.~Heinrich,
  %``Second-order QCD corrections to the thrust distribution,''
  Phys.\ Rev.\ Lett.\  {\bf 99} (2007) 132002
  [arXiv:0707.1285 [hep-ph]].
  %%CITATION = ARXIV:0707.1285;%%
  %122 citations counted in INSPIRE as of 04 juin 2015
  
%\cite{GehrmannDeRidder:2007hr}
\bibitem{GehrmannDeRidder:2007hr}
  A.~Gehrmann-De Ridder, T.~Gehrmann, E.~W.~N.~Glover and G.~Heinrich,
  %``NNLO corrections to event shapes in e+ e- annihilation,''
  JHEP {\bf 0712} (2007) 094
  [arXiv:0711.4711 [hep-ph]].
  %%CITATION = ARXIV:0711.4711;%%
  %155 citations counted in INSPIRE as of 04 juin 2015
  
 %\cite{Dissertori:2007xa}
\bibitem{Dissertori:2007xa}
  G.~Dissertori, A.~Gehrmann-De Ridder, T.~Gehrmann, E.~W.~N.~Glover, G.~Heinrich and H.~Stenzel,
  %``First determination of the strong coupling constant using NNLO predictions for hadronic event shapes in e+ e- annihilations,''
  JHEP {\bf 0802} (2008) 040
  [arXiv:0712.0327 [hep-ph]].
  %%CITATION = ARXIV:0712.0327;%%
  %92 citations counted in INSPIRE as of 04 juin 2015
  
 %\cite{Dissertori:2009ik}
\bibitem{Dissertori:2009ik}
  G.~Dissertori, A.~Gehrmann-De Ridder, T.~Gehrmann, E.~W.~N.~Glover, G.~Heinrich, G.~Luisoni and H.~Stenzel,
  %``Determination of the strong coupling constant using matched NNLO+NLLA predictions for hadronic event shapes in e+e- annihilations,''
  JHEP {\bf 0908} (2009) 036
  [arXiv:0906.3436 [hep-ph]].
  %%CITATION = ARXIV:0906.3436;%%
  %66 citations counted in INSPIRE as of 04 juin 2015
  
  %\cite{Flacher:2008zq}
\bibitem{Flacher:2008zq}
  H.~Flacher, M.~Goebel, J.~Haller, A.~Hocker, K.~Monig and J.~Stelzer,
  %``Revisiting the Global Electroweak Fit of the Standard Model and Beyond with Gfitter,''
  Eur.\ Phys.\ J.\ C {\bf 60} (2009) 543
   [Eur.\ Phys.\ J.\ C {\bf 71} (2011) 1718]
  [arXiv:0811.0009 [hep-ph]].
  %%CITATION = ARXIV:0811.0009;%%
  %226 citations counted in INSPIRE as of 04 juin 2015
  
  %\cite{Glasman:2007sm}
\bibitem{Glasman:2007sm}
  C.~Glasman [H1 and ZEUS Collaborations],
  %``Precision measurements of alpha(s) at HERA,''
  J.\ Phys.\ Conf.\ Ser.\  {\bf 110} (2008) 022013
  [arXiv:0709.4426 [hep-ex]].
  %%CITATION = ARXIV:0709.4426;%%
  %22 citations counted in INSPIRE as of 04 juin 2015
  
%\cite{Alekhin:2012ig}
\bibitem{Alekhin:2012ig}
  S.~Alekhin, J.~Blumlein and S.~Moch,
  %``Parton Distribution Functions and Benchmark Cross Sections at NNLO,''
  Phys.\ Rev.\ D {\bf 86} (2012) 054009
  [arXiv:1202.2281 [hep-ph]].
  %%CITATION = ARXIV:1202.2281;%%
  %207 citations counted in INSPIRE as of 04 juin 2015

%\cite{Martin:2009bu}
\bibitem{Martin:2009bu}
  A.~D.~Martin, W.~J.~Stirling, R.~S.~Thorne and G.~Watt,
  %``Uncertainties on alpha(S) in global PDF analyses and implications for predicted hadronic cross sections,''
  Eur.\ Phys.\ J.\ C {\bf 64} (2009) 653
  [arXiv:0905.3531 [hep-ph]].
  %%CITATION = ARXIV:0905.3531;%%
  %361 citations counted in INSPIRE as of 04 juin 2015
  
%\cite{Ball:2011us}
\bibitem{Ball:2011us}
  R.~D.~Ball, V.~Bertone, L.~Del Debbio, S.~Forte, A.~Guffanti, J.~I.~Latorre, S.~Lionetti and J.~Rojo {\it et al.},
  %``Precision NNLO determination of alpha_s(M_Z) using an unbiased global parton set,''
  Phys.\ Lett.\ B {\bf 707} (2012) 66
  [arXiv:1110.2483 [hep-ph]].
  %%CITATION = ARXIV:1110.2483;%%
  %43 citations counted in INSPIRE as of 04 Jun 2015  
  
 %\cite{Rojo:2014kta}
\bibitem{Rojo:2014kta}
  J.~Rojo,
  %``Constraints on parton distributions and the strong coupling from LHC jet data,''
  arXiv:1410.7728 [hep-ph].
  %%CITATION = ARXIV:1410.7728;%%
  %3 citations counted in INSPIRE as of 04 juin 2015 
  
 %\cite{Aad:2014bia}
\bibitem{Aad:2014bia}
  G.~Aad {\it et al.}  [ATLAS Collaboration],
  %``Jet energy measurement and its systematic uncertainty in proton-proton collisions at $\sqrt{s}=7$ TeV with the ATLAS detector,''
  Eur.\ Phys.\ J.\ C {\bf 75} (2015) 1,  17
  [arXiv:1406.0076 [hep-ex]].
  %%CITATION = ARXIV:1406.0076;%%
  %100 citations counted in INSPIRE as of 04 Jun 2015 
  
%\cite{Chatrchyan:2011ds}
\bibitem{Chatrchyan:2011ds}
  S.~Chatrchyan {\it et al.}  [CMS Collaboration],
  %``Determination of Jet Energy Calibration and Transverse Momentum Resolution in CMS,''
  JINST {\bf 6} (2011) P11002
  [arXiv:1107.4277 [physics.ins-det]].
  %%CITATION = ARXIV:1107.4277;%%
  %526 citations counted in INSPIRE as of 04 juin 2015
  
  %\cite{Chatrchyan:2013txa}
\bibitem{Chatrchyan:2013txa}
  S.~Chatrchyan {\it et al.}  [CMS Collaboration],
  %``Measurement of the ratio of the inclusive 3-jet cross section to the inclusive 2-jet cross section in pp collisions at $\sqrt{s}$ = 7 TeV and first determination of the strong coupling constant in the TeV range,''
  Eur.\ Phys.\ J.\ C {\bf 73} (2013) 10,  2604
  [arXiv:1304.7498 [hep-ex]].
  %%CITATION = ARXIV:1304.7498;%%
  %43 citations counted in INSPIRE as of 04 Jun 2015
  
  \bibitem{ATLAS3jet} 
  ATLAS Collaboration, ATLAS-CONF-2013-041, 2013.

%\cite{Pires:2014rxa}
\bibitem{Pires:2014rxa}
  J.~Pires,
  %``Precise QCD predictions for jet production at the LHC,''
  EPJ Web Conf.\  {\bf 90} (2015) 07005
  [arXiv:1412.3427 [hep-ph]].
  %%CITATION = ARXIV:1412.3427;%%
  %1 citations counted in INSPIRE as of 04 Jun 2015

%\cite{Chatrchyan:2013haa}
\bibitem{Chatrchyan:2013haa}
  S.~Chatrchyan {\it et al.}  [CMS Collaboration],
  %``Determination of the top-quark pole mass and strong coupling constant from the t t-bar production cross section in pp collisions at $\sqrt{s}$ = 7 TeV,''
  Phys.\ Lett.\ B {\bf 728} (2014) 496
   [Phys.\ Lett.\ B {\bf 728} (2014) 526]
  [arXiv:1307.1907 [hep-ex]].
  %%CITATION = ARXIV:1307.1907;%%
  %64 citations counted in INSPIRE as of 04 juin 2015
  
  %\cite{Chatrchyan:2012bra}
\bibitem{Chatrchyan:2012bra}
  S.~Chatrchyan {\it et al.}  [CMS Collaboration],
  %``Measurement of the $t\bar{t}$ production cross section in the dilepton channel in $pp$ collisions at $\sqrt{s}=7$ TeV,''
  JHEP {\bf 1211} (2012) 067
  [arXiv:1208.2671 [hep-ex]].
  %%CITATION = ARXIV:1208.2671;%%
  %116 citations counted in INSPIRE as of 04 juin 2015
  

%\cite{Schmelling:1994pz}
\bibitem{Schmelling:1994pz}
  M.~Schmelling,
  %``Averaging correlated data,''
  Phys.\ Scripta {\bf 51} (1995) 676.
  %%CITATION = PHSTB,51,676;%%
  %29 citations counted in INSPIRE as of 04 juin 2015
  
  

  
%\cite{Hikasa:1992je}
\bibitem{Hikasa:1992je}
  K.~Hikasa {\it et al.}  [Particle Data Group Collaboration],
  %``Review of particle properties. Particle Data Group,''
  Phys.\ Rev.\ D {\bf 45} (1992) S1
   [Phys.\ Rev.\ D {\bf 46} (1992) 5210].
  %%CITATION = PHRVA,D45,S1;%%
  %2030 citations counted in INSPIRE as of 04 juin 2015
  
  %\cite{Montanet:1994xu}
\bibitem{Montanet:1994xu}
  L.~Montanet {\it et al.}  [Particle Data Group Collaboration],
  %``Review of particle properties. Particle Data Group,''
  Phys.\ Rev.\ D {\bf 50} (1994) 1173.
  %%CITATION = PHRVA,D50,1173;%%
  %2826 citations counted in INSPIRE as of 04 juin 2015


%\cite{Schmelling:1996wm}
\bibitem{Schmelling:1996wm}
  M.~Schmelling,
  %``Status of the strong coupling constant,''
  In *Warsaw 1996, ICHEP '96, vol. 1* 91-102
  [hep-ex/9701002].
  %%CITATION = HEP-EX/9701002;%%

%\cite{Bethke:2004uy}
\bibitem{Bethke:2004uy}
  S.~Bethke,
  %``$\alpha_s$ at Zinnowitz 2004,''
  Nucl.\ Phys.\ Proc.\ Suppl.\  {\bf 135} (2004) 345
  [hep-ex/0407021].
  %%CITATION = HEP-EX/0407021;%%
  %115 citations counted in INSPIRE as of 04 Jun 2015
  
  %\cite{Bethke:2006ac}
\bibitem{Bethke:2006ac}
  S.~Bethke,
  %``Experimental tests of asymptotic freedom,''
  Prog.\ Part.\ Nucl.\ Phys.\  {\bf 58} (2013) 351
  [hep-ex/0606035].
  %%CITATION = HEP-EX/0606035;%%
  %228 citations counted in INSPIRE as of 04 juin 2015
  

\end{thebibliography}
\end{document}